\documentclass{WileyMSP-template}
\usepackage{times}
\usepackage{amsmath}
\usepackage{newtxmath}
\usepackage{amsfonts}
\usepackage{graphicx}
\usepackage[version=3]{mhchem} 
\usepackage[detect-weight=true, detect-family=true]{siunitx} 

\usepackage[latin1]{inputenc}
\usepackage{xspace}
\usepackage{color}

\usepackage[hidelinks]{hyperref}
\usepackage{bm}
\usepackage{mathtools}

\begin{document}

\pagestyle{fancy}
\rhead{\includegraphics[width=2.5cm]{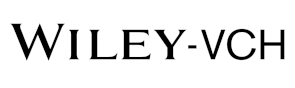}}

\title{Tip-induced nano-engineering of strain, bandgap, and exciton dynamics in 2D semiconductors}

\maketitle


\author{Yeonjeong Koo,}
\author{Yongchul Kim,}
\author{Soo Ho Choi,}
\author{Hyeongwoo Lee,}
\author{Jinseong Choi,}
\author{Dong Yun Lee,}
\author{Mingu Kang,}
\author{Hyun Seok Lee,}
\author{Ki Kang Kim,}
\author{Geunsik Lee, and}
\author{Kyoung-Duck Park$^\ast$}

\begin{affiliations}
Y. Koo, H. Lee, J. Choi, Dr. D. Y. Lee, M. Kang, Prof. K.-D. Park$^\ast$\\
Department of Physics, Ulsan National Institute of Science and Technology (UNIST), Ulsan 44919, Republic of Korea\\
Email Address: kdpark@unist.ac.kr

Y. Kim, Prof. G. Lee\\
Department of Chemistry, Ulsan National Institute of Science and Technology (UNIST), Ulsan 44919, Republic of Korea

Dr. S. H. Choi, Prof. K. K. Kim\\
Center for Integrated Nanostructure Physics (CINAP), Institute for Basic Science (IBS),  Sungkyunkwan University, Suwon, 16419, Republic of Korea

Prof. H. S. Lee\\
Department of Physics, Research Institute for Nanoscale Science and Technology, Chungbuk National University, Cheongju, 28644 Republic of Korea

Prof. G. Lee\\
Center for Wave Energy Materials, Ulsan National Institute of Science and Technology (UNIST), Ulsan 44919, Republic of Korea

Prof. K. K. Kim\\
Department of Energy Science, Sungkyunkwan University, Suwon, 16419, Republic of Korea

\end{affiliations}



\begin{abstract}
\justify
The tunability of the bandgap, absorption and emission energies, photoluminescence (PL) quantum yield, exciton transport, and energy transfer in transition metal dichalcogenide (TMD) monolayers provides a new class of functions for a wide range of ultrathin photonic devices.
Recent strain-engineering approaches have enabled us to tune some of these properties, yet dynamic control at the nanoscale with real-time and -space characterizations remains a challenge.
Here, we demonstrate a dynamic nano-mechanical strain-engineering of naturally-formed wrinkles in a WSe$_2$ monolayer, with real-time investigation of nano-spectroscopic properties using hyperspectral adaptive tip-enhanced PL ($a$-TEPL) spectroscopy.
First, we characterize nanoscale wrinkles through hyperspectral $a$-TEPL nano-imaging with $<$15 nm spatial resolution which reveals the modified nano-excitonic properties by the induced tensile strain at the wrinkle apex, e.g., an increase in the quantum yield due to the exciton funneling, decrease in PL energy up to $\sim$10 meV, and a symmetry change in the TEPL spectra caused by the reconfigured electronic bandstructure.
We then dynamically engineer the local strain by pressing and releasing the wrinkle apex through an atomic force tip control.
This nano-mechanical strain-engineering allows us to tune the exciton dynamics and emission properties at the nanoscale in a reversible fashion.
In addition, we demonstrate a systematic switching and modulation platform of the wrinkle emission, which provides a new strategy for robust, tunable, and ultracompact nano-optical sources in atomically thin semiconductors.

\end{abstract}


\section{Introduction}
\justify
Since the first discovery of emerging photoluminescence (PL) in a direct bandgap MoS$_2$ monolayer \cite{splendiani2010, mak2010}, extraordinarily intriguing physical properties of layered transition metal dichalcogenides (TMDs), such as strong coulomb interaction \cite{van2018}, large spin-orbit coupling \cite{kosmider2013}, and valley-selective dichroism \cite{sie2018}, have been intensively studied for the last decade.
However, despite a broad range of demonstrated platforms for the optoelectronic devices \cite{mak2016}, several hurdles remain before TMD devices can be deployed on a commercial scale.
The presence of nanoscale defects \cite{park2016tmd, park2017}, e.g., grain boundary, edge, and wrinkle, is one of the most significant issues because the 2D nature of the TMD crystals gives rise to complex interactions between atoms and lattices, as well as electrons, phonons, photons, and excitons at these local defects \cite{shi2013}.
Specifically, wrinkles are an inevitable structural deformation in most as-grown and transferred TMD monolayers which gives rise to spatial heterogeneity in material properties \cite{miro2013, luo2015}.
Hence, a comprehensive understanding of wrinkles at their natural length scale is of fundamental significance when exploring the richness of the physics, as well as when trying to greatly improve the performance of optoelectronic devices.

Recently, substantial efforts have been made to reveal the specific nature of nanoscale wrinkles.
Local modifications in lattice, topography, and mechanical properties were understood by transmission electron microscopy (TEM) and atomic force microscopy (AFM) studies \cite{eda2012, deng2017, dagdeviren2020}.
Furthermore, scanning tunneling microscopy (STM) experiments revealed the local electronic density of states and a bandgap reduction at the wrinkle \cite{mills2016, trainer2019}.
However, the optical and excitonic properties of nanoscale wrinkles, i.e., the most intriguing properties for device applications, have not yet been investigated in details due to the diffraction-limited spatial resolution of conventional PL spectroscopy \cite{park2016tmd}.
In addition, a correlated analysis of the structural, electronic, optical, and excitonic properties combined with theoretical simulations is highly desirable to obtain a complete picture of nanoscale wrinkles.

Another prevailing theme in atomically thin semiconductors is strain engineering to achieve bandgap tunable photonic devices \cite{castellanos2013, trainer2019, khan2020, lee2020}.
With large flexibility based on the atomic thickness of 2D TMDs, the induced strain facilitates significant modifications in electronic, optical, and excitonic properties, e.g., electron-phonon and exciton-phonon interactions \cite{christiansen2017, niehues2018}, charge carrier mobility \cite{miro2013}, spin-orbit coupling \cite{koskinen2014}, electronic band structure \cite{niehues2018}, and exciton funneling \cite{lee2020}.
To date, various strain engineering methods have been proposed using substrate deformation devices \cite{chen2017}, indentation of AFM tips \cite{manzeli2015}, piezoelectric actuators \cite{hui2013}, and thermal expansion mismatches \cite{wang2019}.
However, there has been no attempt to control the induced strain of naturally-formed nanoscale wrinkles, while simultaneously investigating their modified nano-optical properties.
On the one hand, a wrinkle is a local defect that causes heterogeneous properties and functions in device applications.
But on the contrary, nano-mechanical engineering of naturally formed wrinkles can be exquisitely employed to develop a controllable nano-optical emitter, which enables tunable bandgap, PL energy, and quantum yield at the nanoscale for quantum optics applications \cite{park2018dark, park2019}.
In addition, most importantly, a wrinkle can act as a nanoscale regulator controlling the exciton funneling and dynamics, e.g., exciton flux and energy conversion, in semiconductors for the realization of highly-efficient optoelectronic devices \cite{yuan2016, high2008, wu2014, unuchek2018, harats2020}.

Here, we present a hyperspectral tip-enhanced photoluminescence (TEPL) nano-imaging approach, combined with nano-optomechanical strain control, to investigate and control the nano-optical and -excitonic properties of naturally-formed wrinkles in a WSe$_2$ monolayer (ML).
In this approach, the excitation field is highly localized at the plasmonic tip by adopting the near-field wavefront shaping technique \cite{lee2020slm} and the tip selectively probes Purcell-enhanced TEPL emission at wrinkles with $<$15 nm spatial resolution.
From hyperspectral TEPL imaging of nanoscale wrinkles, we observe that uniaxial tensile strain locally alters the radiative emission properties, e.g., an increased quantum yield, decreased PL energy up to $\sim$10 meV, and a symmetry change in spectral shape.
Through correlation analyses of the obtained nanoscale maps for topography, TEPL integrated intensity, peak energy, and linewidth, we obtain a complete picture of the exciton funneling behavior at wrinkles.
In addition, we demonstrate the nanoscale strain-engineering of naturally formed wrinkles in a WSe$_2$ ML which allows dynamic control of the nano-excitonic properties in TMDs.
Furthermore, with the ability to reversibly tune the TEPL energy and intensity, we demonstrate switching and modulation of the nano-optical emissions at wrinkles in time and space.

\section{Results and discussion}

\subsection{Hyperspectral $a$-TEPL spectroscopy and tip-induced control}

\begin{figure}
\center
\includegraphics[width=18cm]{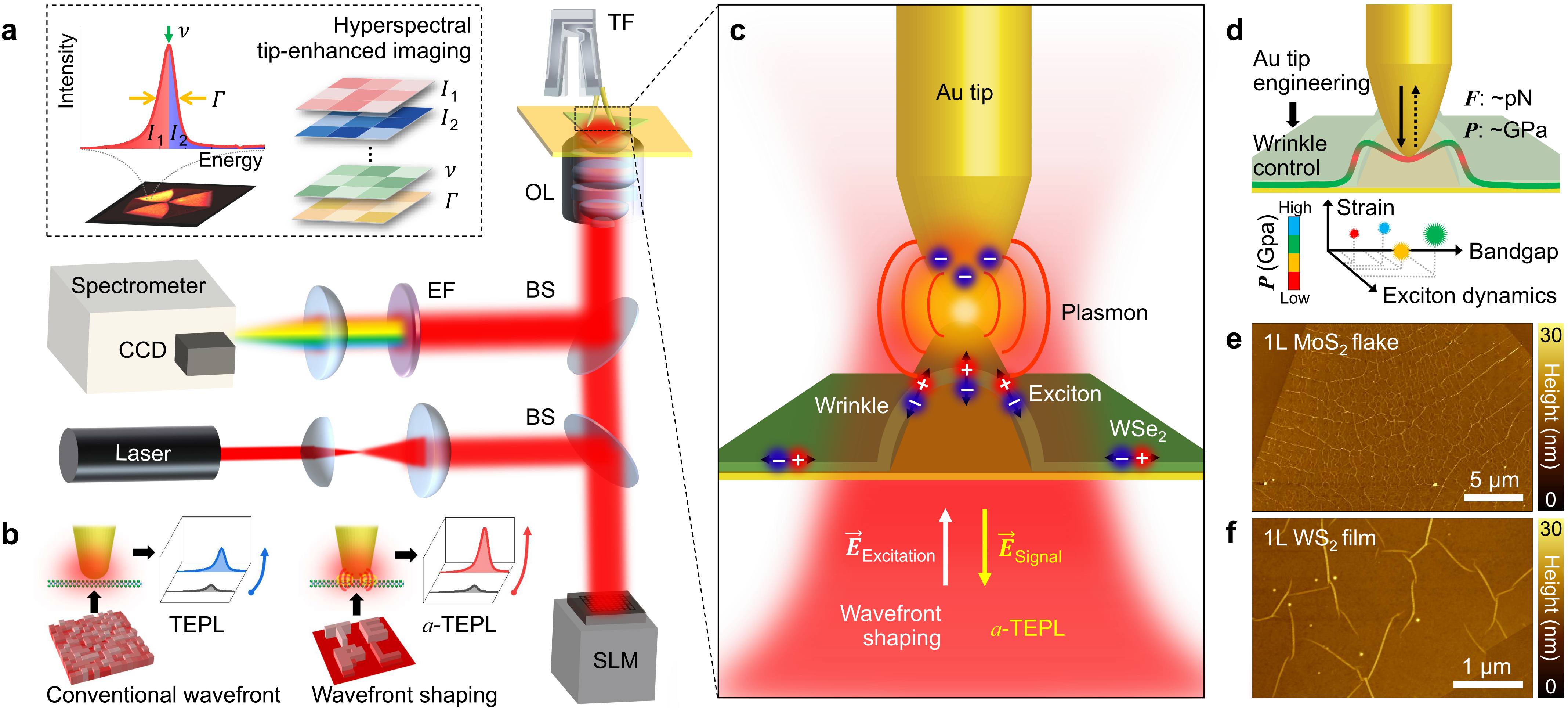}
\caption{
a) Schematic diagram of adaptive TEPL ($a$-TEPL) spectroscopy based on shear-force AFM using bottom-illumination optics with a 632.8 nm He/Ne laser. Inset (top-left): description of the hyperspectral TEPL imaging providing various spatio-spectral information of samples at the nanoscale. 
b) Description of the wavefront shaping effect of $a$-TEPL providing an additional signal enhancement compared to the conventional TEPL using a flat wavefront. 
c) Illustration of plasmon-exciton coupling of a wrinkled WSe$_2$ monolayer at the apex of Au tip giving rise to strong $a$-TEPL responses. 
d) Illustration of the tip-induced nano-engineering of wrinkles to control strain, bandgap, and exciton dynamics.
e-f) AFM topography images of monolayer MoS$_2$ flake and WS$_2$ film transferred onto SiO$_2$ substrate exhibiting naturally-formed nanoscale wrinkles.
Abbreviations: beam splitter (BS), spatial light modulator (SLM), objective lens (OL), tuning fork (TF), and edge filter (EF). 
}
\label{fig1}
\end{figure}

In order to investigate wrinkles in a WSe$_2$ ML, we perform $a$-TEPL experiments based on the bottom-illumination mode tip-enhanced spectroscopy setup with a shear-force AFM using an electrochemically etched Au tip, as shown in Figure \ref{fig1}.
We prepare the wrinkled ML crystal on the thin Au film and excite it with a radially polarized beam.
Furthermore, we use a spatial light modulator (SLM) for wavefront shaping of the excitation beam, which allows us to obtain the additional enhancement of TEPL signals (see Methods for details) \cite{lee2020slm}.
Figure \ref{fig1}B describes a conventional TEPL experiment excited by a flat wavefront (left) and $a$-TEPL experiment excited by the optimal wavefront (right).
In the latter case, the excitation field effectively couples to the tip which in turn, leads to a larger TEPL enhancement compared to the former case.
Specifically, as illustrated in Figure \ref{fig1}C, the coupled light to the Au tip induces resonance oscillation between the optical field and the electron cloud, so called localized surface plasmon resonance (LSPR).
The induced plasmons then interact with the excitons of the WSe$_2$ ML, i.e., plasmon-exciton coupling, which enhances PL response of the crystal due to the increased excitation rate, as well as the Purcell effect \cite{park2018dark}.
Figure 1D describes the tip-induced nano-engineering of wrinkles to manipulate strain, bandgap, and exciton dynamics at the nanoscale.
Since we can regulate the position of Au tip with $\sim$0.2 nm precision in ambient condition \cite{park2018dark}, we can apply the local force and pressure to the nanoscale wrinkle, which enables nanoscale tuning of strain, bandgap, and exciton dynamics in atomically thin semiconductors (see Methods for details and results of Figures \ref{fig4} and \ref{fig5}).
While tip-enhanced nano-spectroscopy has traditionally been regarded as a spectroscopic characterization method, we present an advanced functionality of near-field spectroscopy to control light-matter interactions at the nanoscale.
Figure 1E and F show AFM topography images of nanoscale wrinkles in monolayer MoS$_2$ flake and WS$_2$ film transferred onto SiO$_2$ substrate.
We use a WSe$_2$ monolayer as a model system of this work, but nanoscale wrinkles are unavoidably formed in most layered van der Waals materials transferred on any substrate.

\begin{figure}
\center
\includegraphics[width=17cm]{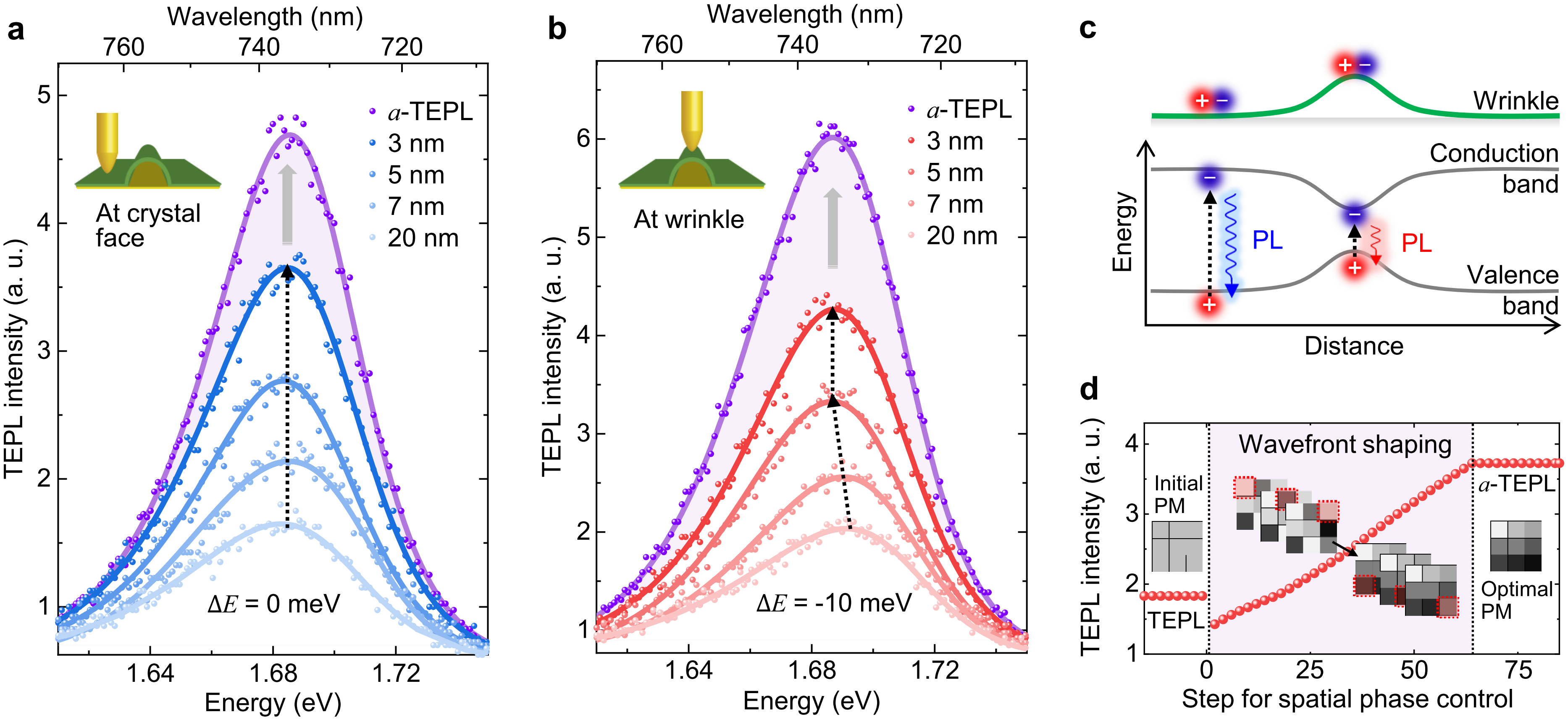}
\caption{
Evolving PL spectra of a WSe$_2$ monolayer as the distance $d$ between the tip and sample is decreased from 20 nm (far-field PL) to 3 nm (TEPL) at the crystal face (a) and wrinkle (b). 
Additionally, $a$-TEPL spectra measured at 3 nm gap are exhibited to demonstrate the wavefront shaping effect. 
c) Schematic illustration for the energy band diagram of a wrinkled WSe$_2$ describing bandgap reduction and corresponding PL redshift due to the naturally induced tensile strain at the wrinkle.
d) Measurement result of evolving $a$-TEPL intensity of the WSe$_2$ crystal during the wavefront shaping. 
}
\label{fig2}
\end{figure}

\subsection{Spectroscopic investigations of exciton properties of nanoscale wrinkles}

To understand the modified excitonic properties of wrinkles in a WSe$_2$ ML, we investigated the nanospectroscopic properties of exciton emissions at the wrinkle and crystal face by measuring the TEPL spectra with respect to the distance ($d$) between the tip and sample.
Figure \ref{fig2}A and B show the representative far-field and TEPL spectra (far-field PL at $d$ = 20 nm and TEPL at $d$ = 7, 5, and 3 nm) acquired at the crystal face and the wrinkle. 
At the crystal face, PL intensity increases as the tip approaches the crystal through the plasmon-exciton coupling, yet the PL energy and spectral shape are invariant (A, blue).
By contrast, the distance-dependent TEPL spectra at the wrinkle show gradual PL energy shift up to -10 meV with PL enhancement (B, red).
At $d$ = 20 nm, we observe a similar PL spectrum to the TEPL spectra measured at the crystal face because the dominant far-field PL signal from the diffraction-limited beam spot cannot resolve the local excitonic properties of the nanoscale wrinkle.
At $d$ = 7 nm, the enhanced local near-field PL response starts to emerge with the PL redshift, but the contribution of the far-field PL response is still significant.
The influence of the far-field PL response is negligible from $d$ = 5 nm, i.e., the peak position of the TEPL spectra is not further redshifted at smaller distances than $d$ = 5 nm, and TEPL intensity is further enhanced as the tip closely approaches the wrinkle apex ($d$ = 3 nm).
The observed PL redshift at the wrinkle is associated with the modified electronic bandstructure due to the induced uniaxial tensile strain at the apex of wrinkles, as described in Figure \ref{fig2}C.
The bandgap reduction and PL energy shift at the nanoscale structure are in good agreement with the previous reports on the microscale wrinkled structures \cite{castellanos2013, trainer2019, khan2020, lee2020}.

From this measurement, we quantified the TEPL enhancement factor (EF) at $d$ = 3 nm as high as $\sim$4.7$\times$$10^3$ (see Section S1 for details), attributed to the enhanced excitation rate (${\left| {E_{NF}}/{E_{FF}} \right|}^{2}$) as well as the Purcell factor (${\gamma}_{PF}={\gamma}/{\gamma}_0$) given by
\begin{equation}\label{eq1}
<\text{TEPL EF}> = {\left| {E_{NF}}\over{E_{FF}} \right|}^2\times \gamma_{PF},
\end{equation}

\noindent where $E_{NF}$ and $E_{FF}$ are optical field amplitudes of near-field and far-field excitation beams and $\gamma/{\gamma}_0$ is the enhanced spontaneous emission rate in the plasmonic cavity \cite{park2018dark}. 
To obtain additional enhancement of TEPL signal, we adaptively control the excitation light using a SLM, i.e., we switch the excitation beam from the conventional flat wavefront to the optimal wavefront customized for the Au tip, as shown in Figure \ref{fig2}A and B (see Fig. S2 for the used phase masks).
The optimal phase mask of SLM was found via the stepwise sequential algorithm before we perform the distance-dependence TEPL experiment.
Once we find the optimal wavefront of the specific Au tip, the wavefront switching can be performed dynamically with a temporal resolution of $\sim$20 ms \cite{lee2020slm}.
Thus, we obtain $a$-TEPL spectra at $d$ = 3 nm in Figure \ref{fig2}A and B (purple) by immediately switching the wavefront without affecting to shear-force feedback conditions.
Through the wavefront shaping, we achieve additional TEPL enhancement of $\sim$130 \%, which allows TEPL nano-spectroscopy and -imaging experiments with higher signal to noise ratio.
Note that the $a$-TEPL enhancement factor is $\sim$6.1$\times$$10^3$, which is a $\sim$30 \% higher value than the TEPL enhancement factor because far-field PL intensity is almost unchanged by the wavefront shaping.

Figure \ref{fig2}D shows a change in the measured $a$-TEPL intensity of a WSe$_2$ ML during the wavefront shaping process for the different Au tip, which gives $\sim$200 \% larger additional enhancement compared to normal TEPL intensity (see Section S3 for details), with illustration to find the optimal phase mask.
As we described in detail in the Method section, we set a random phase first for the 64 segments to prevent the local minimum problem \cite{lee2020slm} and find the optimal temporal phase of each segment by monitoring $a$-TEPL intensity.
By repeating this feedback process sequentially for the 64 segments, we can find the optimal phase mask giving the maximum $a$-TEPL response.
It should be noted that the optimal phase mask differs from tip-to-tip since the tip apex shape is always slightly different.
A large number of experiments have verified that the $a$-TEPL signal gives $\sim$120-250 \% larger intensity compared to conventional TEPL signals and we found a tendency that sharp and symmetric shape tips have relatively small additional enhancements by the wavefront shaping.

\begin{figure}[!t]
\center
\includegraphics[width=18.5cm]{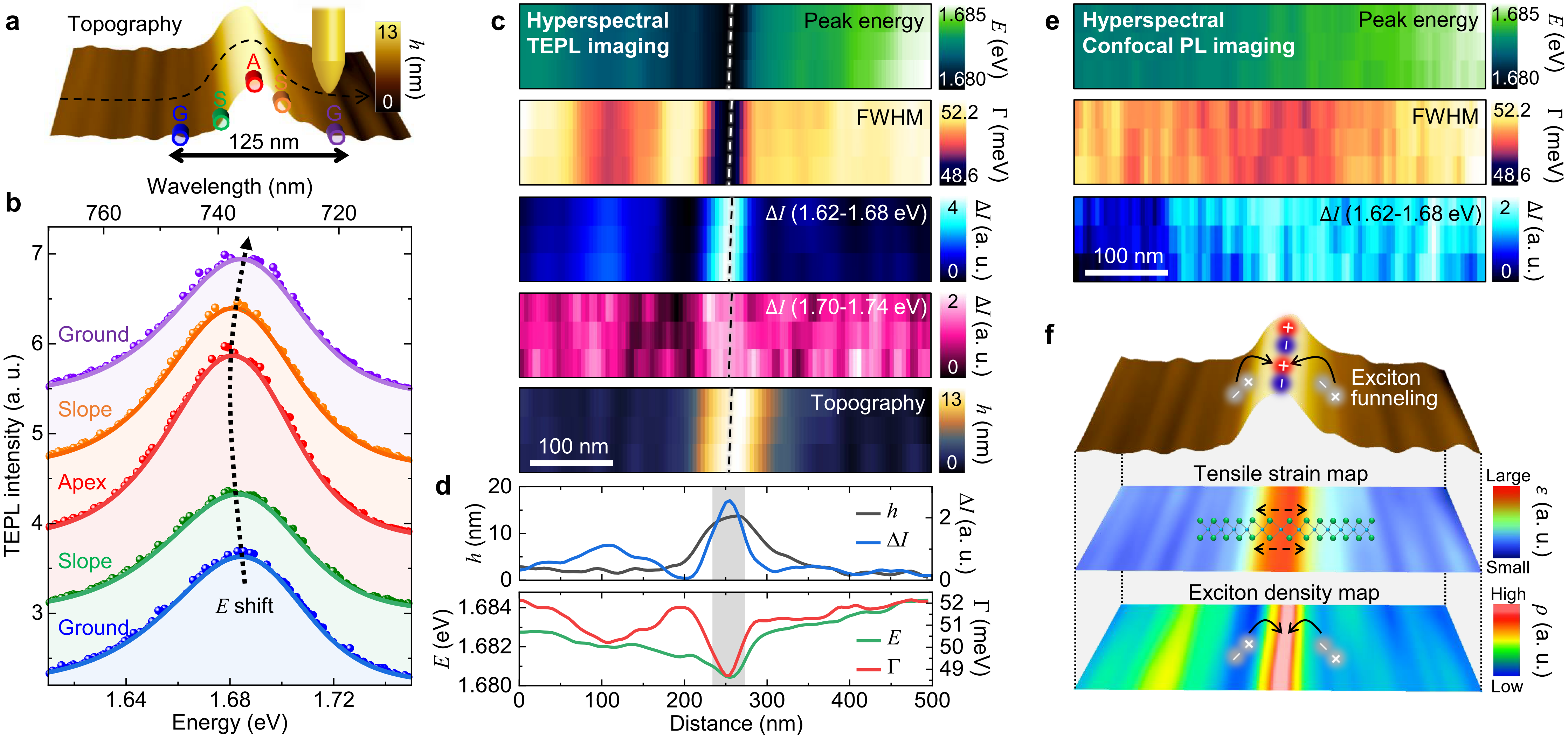}
\caption{
a) Topography of a wrinkle in WSe$_2$ crystal measured with a shear-force AFM. 
b) TEPL spectra measured at the ground (G), slope (S), and apex (A) of the wrinkle indicated in (a), resolving PL characteristics in sub-wavelength scale. 
c) Hyperspectral TEPL images and corresponding AFM topography image of a wrinkle. Peak position ($E$), FWHM (linewidth $\Gamma$), and spectrally integrated intensity ($\Delta$$I$ = 1.62 - 1.68 eV displaying PL redshifted region and $\Delta$$I$ = 1.70 - 1.74 eV displaying uninfluenced spectral region of PL redshift) maps show distinct variations at the wrinkle. The measurement area is 500 nm $\times$ 10 nm with a step of 5 nm. 
d) Line traces of structural and spectral information revealing the excitonic properties of the wrinkle at the nanoscale. 
e) Hyperspectral confocal PL images for the same wrinkle which cannot resolve the optical properties of the wrinkle due to the limited spatial resolution. 
f) Illustration describing the exciton funneling phenomenon into the apex of the wrinkle from the slope regions. 
The expected tensile strain and exciton density maps near the wrinkle based on the correlated analysis of hyperspectral TEPL imaging.
}
\label{fig3}
\end{figure}

\subsection{Spatio-spectroscopic investigations of excitonic properties of nanoscale wrinkles}

To form a more comprehensive picture of the naturally-formed nanoscale wrinkles with the correlation analysis of structural, optical, and excitonic properties, we performed hyperspectral TEPL nano-imaging.
Figure \ref{fig3}A shows the measured topography with the indications of ground (G), slope (S), and apex (A) regions of the wrinkle.
Correspondingly, Figure \ref{fig3}B shows the observed TEPL spectra acquired at these regions.
As the tip ascends to the apex (A) from the ground (G) passing through the slope (S) and then descends back to the ground (G), the peak position of the TEPL spectra is gradually redshifted and then came back to its original energy state.
In addition, the TEPL intensity and spectral shape show distinct variations depending on the location. 
The apex region shows larger intensity with more symmetric spectral shape compared to the TEPL signals measured at the ground regions.
The TEPL spectra measured at the slope regions show intermediate spectral shape with different behaviors in emission intensity, i.e., the larger emission intensity is observed at the slope near the apex (orange) while the smaller intensity is observed at the slope near the ground (green) compared to the TEPL intensity measured at the ground regions (blue, purple). 

The detailed heterogeneous properties are revealed from hyperspectral TEPL nano-images, as shown in Figure \ref{fig3}C.
The TEPL peak energy ($E$) map shows the gradual decrease of the PL energy near the wrinkle and minimum energy of $\sim$1.680 eV at the apex due to the tensile strain.
Similarly, the TEPL linewidth (FWHM) decreases at the wrinkle from $\sim$52 meV to $\sim$48 meV due to the symmetrically modified spectral shape, as shown in Figure \ref{fig3}B.
The asymmetric nature of the TEPL spectrum at the crystal face is attributed to the trion and biexciton shoulders \cite{he2019, park2016tmd} and asymmetric phonon sidebands caused by the strong exciton-phonon coupling \cite{niehues2018}.
Therefore, from the modified spectral shape and the linewidth map, we believe that the exciton PL response is more significantly influenced by the induced tensile strain at the wrinkle compared to the PL responses of trions and biexcitons.
In addition, the coupling between excitons and phonons decreases at the strained region which is a characteristic of the selenium-based TMD monolayers \cite{niehues2018}. 
The spatial variance of the emission intensity is clearly visualized in the spectrally integrated TEPL intensity maps.
The intensity map of the PL redshifted region (integrated intensity $\Delta$$I$ in the spectral region of 1.62 - 1.68 eV, blue) shows sharp change at the wrinkle, while the intensity map of the higher energy area (integrated intensity $\Delta$$I$ in the spectral region of 1.70 - 1.74 eV, pink) shows no distinct variation.
These features are more pronounced in the line traces of the hyperspectral maps, as shown in Figure \ref{fig3}D.
TEPL intensity at the junction between the ground and the slope is lower than TEPL intensity measured at the ground.
By contrast, there is a large increase in TEPL intensity at the apex, even though tensile strain is induced.
This result can be explained by the exciton funneling effect since the photogenerated excitons tends to drift to the lower bandgap region \cite{feng2012, castellanos2013}.
In addition to the physical movement of excitons, i.e., funneling, we need to take into account the energy transfer of excitons to the strained region, i.e., exciton diffusion.
The diffusion length $L$ for excitons in 2D materials is given by $L$ = $2\sqrt{D_{exc}\tau}$, where $D_{exc}$ is the exciton diffusivity and $\tau$ is the exciton lifetime \cite{zande2013}. 
Since the exciton diffusion length of a WSe$_2$ monolayer is several tens of nanometers at room temperature \cite{park2018dark, zande2013}, this energy transfer can affect to the observed TEPL intensity distribution at the nanoscale wrinkle.
We believe these combined effects give rise to bright nano-emitting spot at the apex even though the induced strain is quite small, which is a distinguishing property of the nanoscale wrinkle compared to the previous strain-engineering studies at the microscale \cite{castellanos2013}.
Note that we observed similar features at different wrinkles, but the redshifted PL energy differs from wrinkle to wrinkle (see Section S4 for details).
This variation of wrinkles is associated with the geometric structure of a wrinkle. The induced tensile strain ${\varepsilon}_{\text{tensile}}$ at the wrinkle apex is estimated as \cite{park2018dark}
\begin{equation}\label{eq2}
{\varepsilon}_{\text{tensile}} \cong  {{\pi}^2 h\delta \over{\left({1-{\sigma}^2}\right){\lambda}^2}}, 
\end{equation}

\noindent where $h$ is the thickness of the crystal, $\delta$ and $\lambda$ are the height and width of the wrinkle, and $\sigma$ is the Poisson's ratio of WSe$_2$ (0.19) \cite{zhang2016}. 
Hence, the induced ${\varepsilon}_{\text{tensile}}$ and corresponding PL energy shift are variable for different wrinkles.
It should be also noted that these modified excitonic properties at the nanoscale wrinkles cannot be observed in hyperspectral confocal PL imaging due to the limited spatial resolution (Figure \ref{fig3}E).
Lastly, from correlated information of the hyperspectral TEPL maps, we can illustrate sketchy maps of tensile strain and exciton density at the wrinkle, as shown in Figure \ref{fig3}F. 
These nanoscale pictures of strain, energy shift, and exciton confinement provide better understanding of the naturally-formed wrinkles existing in most transferred TMD crystals.

\begin{figure}
\center
\includegraphics[width=17cm]{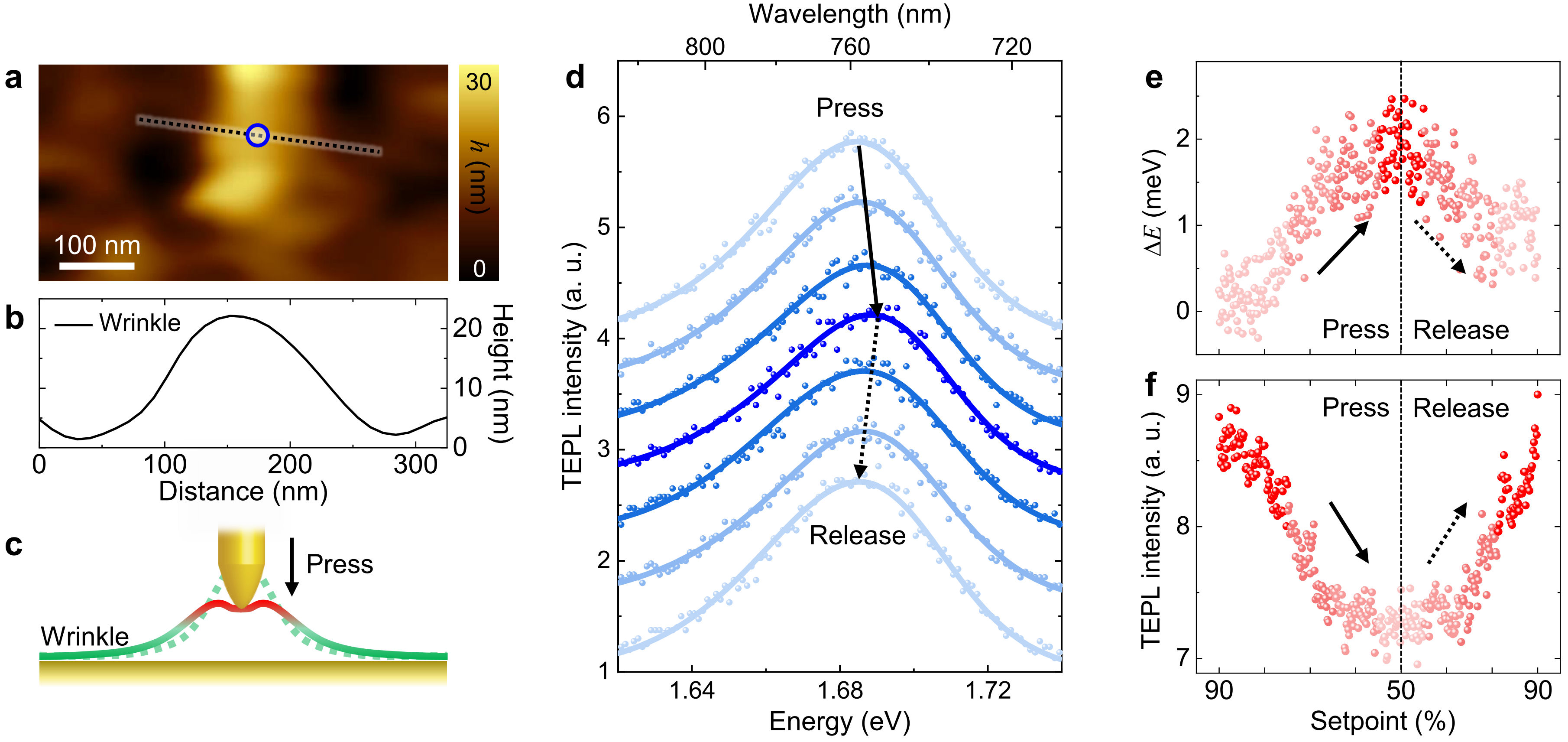}
\caption{
Topography image (a) and line profile (b) of a wrinkle structure in a WSe$_2$ monolayer. 
The line profile is derived from the dotted line indicated in (a). 
Blue circle indicates the pressing point by the tip. 
c) Illustration of the nano-optomechanical control of a wrinkle through the AFM tip-force engineering. Pressing and releasing the wrinkle by the Au tip enables a reversible control of the tensile strain, bandgap, and PL energy at the nanoscale. 
d) TEPL spectra of the wrinkle measured during gradual press and release of the tip, exhibiting a reversible PL energy shift by the nanoscale strain control. 
Peak energy shift (e) and peak intensity change (f) for the TEPL spectra measured during the pressing and releasing the wrinkle with a depth of $\sim$10 nm. 
The depth of the tip in the pressing and releasing process is regulated by the setpoint control in shear-force AFM ranging from 90 \% to 50 \% of the initial value. 
}
\label{fig4}
\end{figure}

\subsection{Tip-induced nano-engineering of nanoscale wrinkles}

A naturally-formed wrinkle can act as a high quantum yield nano-emitter for various photonic device applications.
Furthermore, its application range will be greatly expanded if the nano-optical properties of the wrinkle are dynamically controllable.
Since the recently developed strain engineering approaches are not applicable to the nanoscale wrinkles \cite{castellanos2013, trainer2019, khan2020, lee2020}, we present a nano-optomechanical strain-engineering approach using the Au tip.
We have found a wrinkle structure with a height of $\sim$20 nm and obtained an AFM topography image, as shown in Figure \ref{fig4}A and B.
For strain-engineering of the wrinkle with simultaneous investigation of the modifying excitonic properties, we measured the TEPL spectra while applying a contact force to the wrinkle apex using the Au tip, as illustrated in Figure \ref{fig4}C.
The pressing depth $d_p$ was precisely regulated by changing the setpoint in shear-force feedback (see Methods for details).
Figure \ref{fig4}D shows the selected TEPL spectra measured during gradual press and release of the wrinkle by the tip, exhibiting an emission intensity change with a spectral shift.
As the tip applies a force to the apex, the wrinkle structure is flattened and the induced uniaxial ${\varepsilon}_{\text{tensile}}$ is released, which leads to the PL blueshift with a decreased emission rate due to the modified electronic bandstructure and the suppressed exciton funneling.
As the tip releases, the PL energy is redshifted again with the increasing TEPL intensity attributed to the re-induced ${\varepsilon}_{\text{tensile}}$ at the restored apex structure.
Figure \ref{fig4}E and F show the changing peak energy ($\Delta$$E$) and emission intensity derived from the full TEPL spectra set measured during the process of tip-press and -release, demonstrating sophisticated controllability of the nanoscale wrinkle.
Note that the pressing depth $d_p$ was $\sim$10 nm when we reduced the setpoint to 50 \% compared to the amplitude of tuning fork/tip assembly in free space.
This behavior is similarly observed from other wrinkles reproducibly.
Through this nano-optomechanical tip-interaction control, we can engineer the strain, exciton dynamics, and bandgap energy at the nanoscale in a reversible fashion, which provides a new strategy for creating nano-optical sources in atomically thin semiconductors.

\begin{figure}[!t]
\center
\includegraphics[width=17cm]{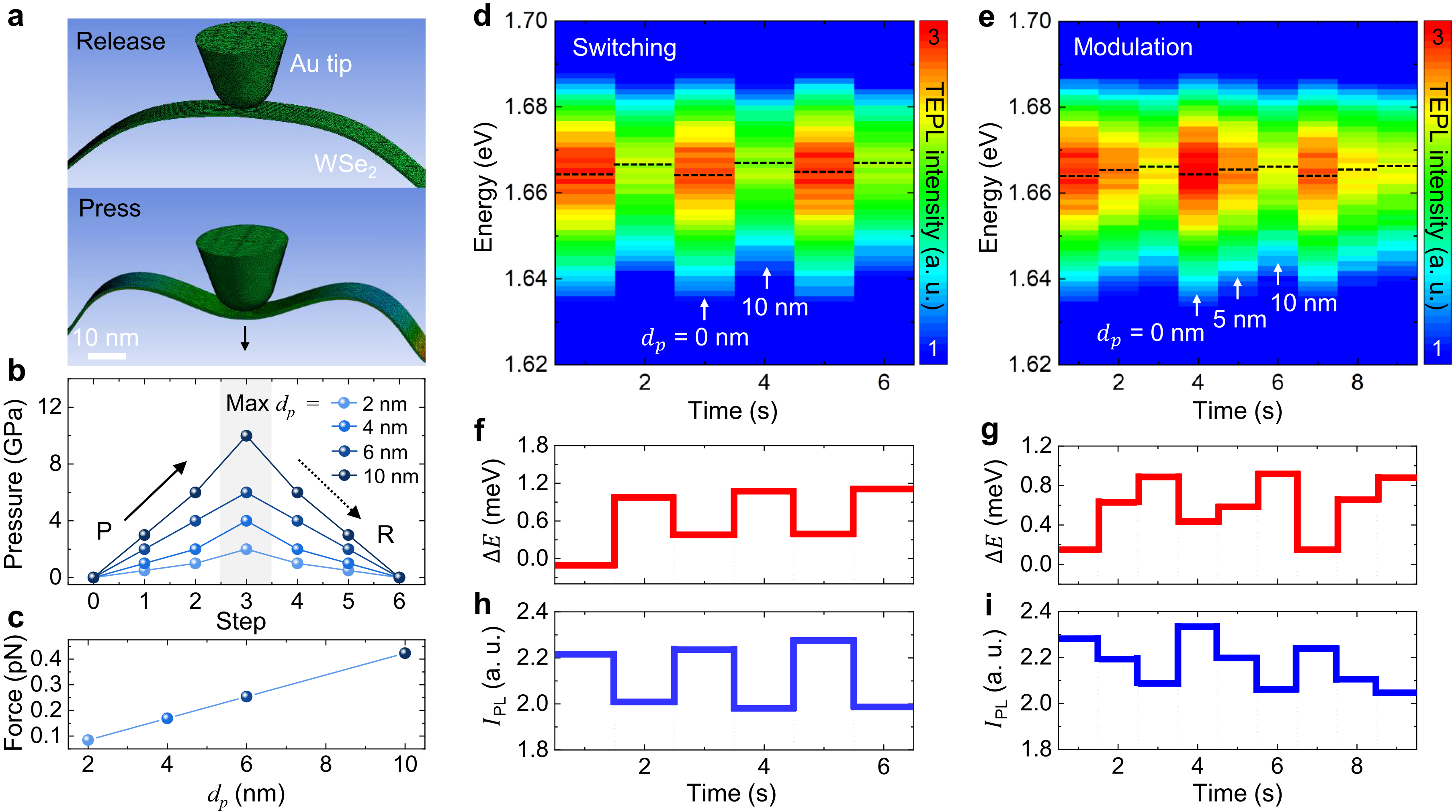}
\caption{
a) Model of the Au tip and the WSe$_2$ wrinkle to simulate the tip-induced local pressure at the wrinkle apex. 
Calculated local pressure (b) and force (c) applied to the wrinkle apex when the Au tip presses and releases for different pressing depths ($d_p$ = 2, 4, 6, and 10 nm). 
d-e) Time-series TEPL responses of a wrinkle with the nanoscale strain engineering by the tip interaction control. 
d) Demonstration of a switching mode for the emission energy and intensity with binary pressing depth of the tip ($d_p$ = 0 nm for bright and lower energy emission and $d_p$ = 10 nm for dark and higher energy emission). 
e) Demonstration of a modulation mode with three discrete levels ($d_p$ = 0, 5, and 10 nm). 
Black dotted lines indicate peak energy of TEPL responses during switching and modulation of the wrinkle emission. 
Corresponding time-series of peak energy shift (f, g) and TEPL intensity ($I_{\text{PL}}$) change (h, i), derived from TEPL spectra of (d) and (e).
}
\label{fig5}
\end{figure}

As we presented in Figure \ref{fig4} and as in previous studies \cite{park2016tmd, park2018dark}, we can precisely control the tip-sample interaction by using a shear-force feedback mechanism.
To quantify the tip-induced local pressure and force at the apex of the wrinkle, we modeled the Au tip and the WSe$_2$ wrinkle using a commercially available numerical simulator (ANSYS), as shown in Figure \ref{fig5}A (see Methods for detailed simulation conditions).
When the tip-pressure is applied to the wrinkle in the area of tip apex, the ${\varepsilon}_{\text{tensile}}$ at the wrinkle is released spatially extending over several tens of nanometers so that the wrinkle shape is flattened (see Movie S1 for simulation process).
To explicitly simulate our experiment in Figure \ref{fig4}, we calculated the local pressure induced at the wrinkle when the tip presses and releases the apex for different pressing depths ($d_p$ = 2, 4, 6, and 10 nm).
Figure \ref{fig5}B shows the calculated maximum pressure at the tip apex region ranging from $\sim$2 GPa to $\sim$10 GPa depending on the $d_p$.
We also calculated the tip-induced local force at different maximum pressing depths, estimated as high as $\sim$0.4 pN, as shown in Figure \ref{fig5}C.
Under exquisite control of the pressing depth, we can precisely manipulate the strain on the wrinkle in a reversible fashion, which in turn means that we can dynamically engineer the radiative emission of nanoscale wrinkles using the Au tip.

To demonstrate the exemplary platform for potential applications in nano-optoelectronic devices, we present switching and modulation modes of the wrinkle emission. 
Figure \ref{fig5}D and E show time-series TEPL responses of a nanoscale wrinkle with a systematic strain-engineering using the tip.
For the switching mode, we use a binary pressing depth of $d_p$ = 0 nm and 10 nm, alternating between a bright and lower energy emission and a dark and higher energy emission (See Methods for details).
For the modulation mode, we simply modify the program code to change the pressing depth to one of three discrete levels ($d_p$ = 0, 5, and 10 nm).
The discrete changes in the emission energy and intensity as a function of time are clearly displayed in Figure \ref{fig5}F-I.
In this nanomechanical photonic device platform, the frequency of switching and modulation can be achieved as high as $\sim$1 kHz determined by the settling time of a quartz tuning fork sensor in a shear-force feedback \cite{karrai1995}.

\subsection{Theoretical quantification of the induced tensile strain at nanoscale wrinkles}

\begin{figure}[!t]
\center
\includegraphics[width=12.5cm]{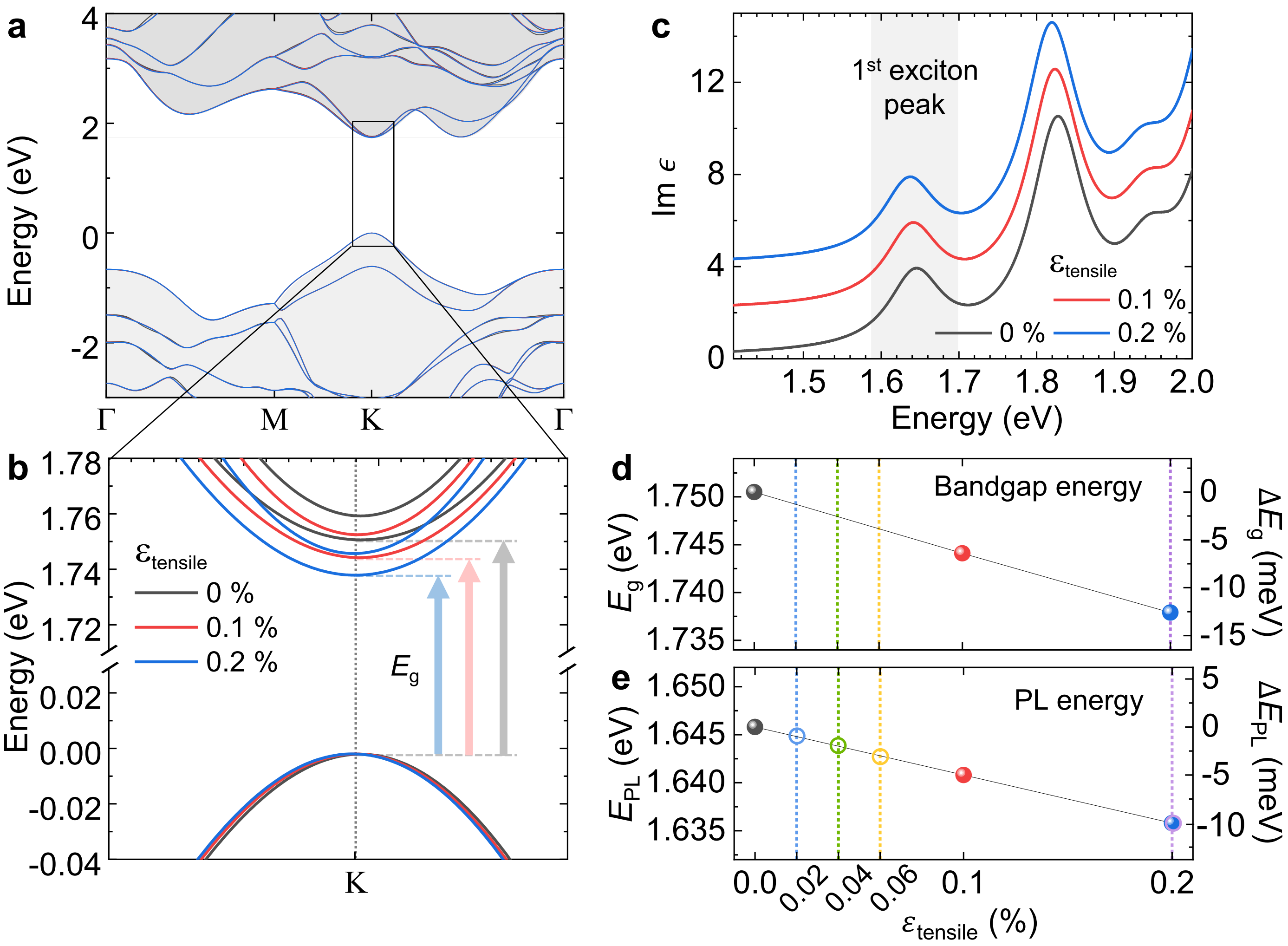}
\caption{
a) HSE06 band structures of monolayer WSe$_2$ at lattice constant of 3.282 \AA{ }(0 \%), 3.285 \AA{ }(0.1 \%), and 3.288 \AA{ }(0.2 \%). 
b) Direct minimum energy gap of three slab structure at K-point. 
c) Imaginary part of the dielectric function $\epsilon$ calculated with the model BSE method. 
d) HSE06 bandgap and e) 1st exciton peak energy versus strain.
}
\label{fig6}
\end{figure}

In order to quantify the induced ${\varepsilon}_{\text{tensile}}$ at the apex of experimentally observed wrinkles, we simulated electronic bandstructures of three different WSe$_2$ slab structures with ${\varepsilon}_{\text{tensile}}$ = 0 (pristine), 0.1, and 0.2 \% using the Vienna ${ab}$ ${initio}$ simulation package (VASP) \cite{kresse1996}, as shown in Figure \ref{fig6}A (See Methods for simulation conditions). 
The density functional theory (DFT) calculation with considering the spin-orbit coupling shows decreasing direct bandgap energies ($E_g$) of 1.750, 1.744, and 1.738 eV at K-point with increasing tensile strain (Figure \ref{fig6}B).
We then simulated the optical spectra based on the Bethe-Salpeter equation (BSE) method using the screening parameters from PBE and PBE+GW calculations \cite{bokdam2016, shishkin2006}. 
Figure \ref{fig6}C shows the calculated optical spectra (imaginary part of the dielectric function $\epsilon$) with exciton peaks, exhibiting the absorbance decreases as the uniaxial strain increases. 
The calculated bandgap energy and absorption resonance of the 1st exciton peak as a function of the tensile strain are displayed in Figure \ref{fig6}D and E. 
Here, the absorption resonance of the 1st exciton peak directly corresponds to the optical energy gap and the PL energy. 
From the obtained values of PL energy shift $\Delta E_{\text{PL}}$ in Figure \ref{fig6}E (0, -5, and -10 meV), we derive a linearly proportional relation of
\begin{equation}\label{eq3}
\Delta E_{\text{PL}} \cong \alpha\cdot{\varepsilon}_{\text{tensile}},
\end{equation}

\noindent where $\alpha$ is a proportionality constant with a value of $\alpha$ = -50 meV$\cdot{\%}^{-1}$. 
Using this equation, we quantify the induced tensile strain of the observed wrinkles in Figures \ref{fig2}, \ref{fig3}, \ref{fig4}, and \ref{fig5}, i.e., ${\varepsilon}_{\text{tensile}}$ $\sim$ 0.2, 0.06, 0.04, and 0.02 \% for the Figures \ref{fig2}B, \ref{fig3}C, \ref{fig4}D, and \ref{fig5}D-E, respectively.
Note that there is a quantitative difference between the theoretically derived values (i.e., bandgap and PL energy) and the experimentally observed values because the simulations only consider a single electron-hole pair even though there are a number of excitons in the used crystal which have complex interactions with lattice, atoms, electrons, and so on.

\section{Conclusions}

In summary, we investigated the correlated structural and optical properties of naturally-formed nanoscale wrinkles in a WSe$_2$ monolayer using hyperspectral $a$-TEPL nano-spectroscopy and -imaging.
This experiment allowed us to reveal the modified electronic properties and exciton behaviors at the wrinkle, associated with the induced uniaxial tensile strain at the apex.
Based on this comprehensive understanding of the nano-optomechanical properties of a nanoscale wrinkle, we exploited the wrinkle structure as a nanoscale strain-engineering platform. 
The precise atomic force tip control enabled to engineer the excitonic properties of TMD monolayers at the nano-local regions in a fully reversible fashion.
Furthermore, we presented a more systematic platform for dynamic nano-emission control of the wrinkle by demonstrating programmably operational switching and modulation modes in time and space.
Hence, we envision that our approach gives access to potential applications in quantum-nanophotonic devices, such as bright nano-optical sources for light-emitting diodes, nano-optical switch/multiplexer for optical integrated circuits, and exciton condensate devices.


\section{Experimental Section}
\threesubsection{Growth of monolayer WSe$_2$ flakes}\\
The tungsten precursor was prepared by dissolving 0.01 M sodium tungstate dihydrate (STD, Na$_2$WO$_4$ $\cdot$ 2H$_2$O, $\ge$ 99 \%, Sigma-Aldrich) in aqueous solution. 
STD solution was spun onto the SiO$_2$/Si substrate with 2500 rpm for 1 min. 
To grow monolayer WSe$_2$ flakes, Se pellets ($\ge$ 99.99 \%, Sigma-Aldrich) and the STD-coated substrate were loaded into the respective upstream and downstream zone in a two-zone furnace system equipped with a 2-inch quartz tube. Prior to growth, the quartz tube was purged by flowing high purity (99.999 \%) Ar with a flow rate of 550 sccm for 10 mins. 
The temperatures of the two heating zones were elevated to $365\,^{\circ}\mathrm{C}$ and $780\,^{\circ}\mathrm{C}$ for 10 mins and maintained for 15 mins. 
After finishing the growth, the quartz tube was naturally cooled down to room temperature. The entire growth procedure was conducted under Ar and H$_2$ atmosphere with flow rates of 550 sccm and 8 sccm, respectively, at atmospheric pressure.\\

\noindent  \threesubsection{Transfer of WSe$_2$ flakes}\\
Cover glasses with a thickness of 170 $\mu$m were ultrasonicated in acetone and isopropanol for 10 mins each and cleaned again with an O$_2$ plasma treatment for 10 mins.
The glasses were then deposited with a Cr adhesion layer with a thickness of 2 nm (a rate of 0.1 \AA/s) and subsequently deposited with an Au film a thickness of 9 nm (a rate of 1.0 \AA/s) at a base pressure of $\sim$10-6 torr, using a conventional thermal evaporator.
The deposition rate of metals was precisely controlled using a quartz crystal microbalance (QCM) detector.
Then, monolayer WSe$_2$ flakes were transferred onto the Au film/cover glass by the conventional wet-transfer method \cite{choi2017}. 
The poly(methyl methacrylate) (PMMA) solution as a supporting layer was spun on WSe$_2$ flakes grown SiO$_2$/Si substrate. 
To delaminate WSe$_2$ from the SiO$_2$/Si substrate, the PMMA-coated sample was floated on 2M KOH aqueous solution. 
After the delamination, the residual KOH underneath the PMMA/WSe$_2$ layer was rinsed by floating on fresh deionized water. 
The PMMA/WSe$_2$ layer was finally transferred onto the gold substrate by scooping method and dried in $90\,^{\circ}\mathrm{C}$ oven for 5 mins.
During these processes, natural wrinkles are always formed. 
The PMMA layer was removed by washing with acetone and isopropyl alcohol.\\

\noindent  \threesubsection{Hyperspectral TEPL nano-imaging setup}\\
For hyperspectral TEPL nano-imaging experiment, we constructed tip-enhanced nano-spectroscopy with the bottom illumination optics based on a shear-force AFM using the Au tip.
The Au tip (apex radius of $\sim$15 nm) was prepared using the refined electrochemical etching protocol \cite{neacsu2005} and attached to a quartz tuning fork with super glue. 
The mechanically dithered tuning fork/Au tip assembly vibrated at its resonance frequency ($\sim$32 kHz) with a Q-factor of $\sim$1000 and the changing resonance properties by shear-force were monitored for tip-sample distance regulation \cite{park2018dark}.
The vertical and lateral positioning of the tip and the lateral positioning of the sample scanner were operated by a digital AFM controller (Solver next SPM controller, NT-MDT).

For hyperspectral spectroscopy and imaging, the excitation and detection optics were built as follows:
To make a spatially coherent excitation beam, He-Ne laser ($\lambda$ = 632.8 nm, optical power $P$ of $\le$0.6 mW) was passed through a single-mode fiber (core diameter of $\le$3.5 $\mu$m) and collimated to a 25 mm diameter beam using an aspheric lens.
The collimated beam was then passed through a half wave plate ($\lambda$/2) and a polarizing beam splitter (PBS) so that the SLM (P512/PDM512, Meadowlark optics) received only the horizontally polarized beam.
The beam reflected from the active area of the SLM, was followed by a 4f system and focused at the Au tip-Au film junction by an oil-immersion objective lens (PLN100x, 1.25 NA, Olympus).
Note that, in between the SLM and the objective lens, we placed a radial polarizer to obtain the vertically polarized beam (in parallel with the optical axis) as large as possible at the focus. 
This leads to a large field enhancement and effective plasmon-exciton coupling at the tip apex which eventually gives highly enhanced TEPL signals.
The backscattered TEPL signals from a sample were collected by the same objective lens and passed through a pinhole (hole diameter of 5 $\mu$m) between two focusing lenses to suppress the far-field PL signal, i.e., the background noise in the TEPL measurement, by adopting a confocal microscopy scheme.
TEPL signals were then sent to a spectrometer (f = 320 mm, Monora320i, Dongwoo Optron) after blocking out the laser line with an edge filter (633 nm cut-off).
The dispersed signals were finally imaged onto a thermoelectrically cooled charge-coupled device (CCD, DU971-BV, Andor) to obtain the TEPL spectra.
The spectrometer was calibrated using an Hg-Ar lamp for a 150 g/mm grating and its spectral resolution of $\sim$1.6 nm was identified.

For hyperspectral mapping, we recorded the TEPL spectra at each pixel during an AFM scanning.
We then performed line fitting for all the recorded TEPL spectra with an asymmetric double sigmoidal function, which gave the best fitting condition for the asymmetric PL spectra of the WSe$_2$ ML.
With the derived fitting parameters, we obtained nano-maps for the integrated TEPL intensity, peak position, and linewidth, as shown in Figure \ref{fig3}C.\\

\noindent \threesubsection{Wavefront shaping for $a$-TEPL}\\
To obtain an additional signal enhancement from the conventional TEPL signals, we used the adaptive optics technique, i.e., wavefront shaping of the excitation beam, we recently developed \cite{lee2020slm}.
Basically, since the apex structure of an electrochemically etched tips differs from tip-to-tip, the adaptive wavefront of $a$-TEPL gives rise to a larger field enhancement than that of a conventional tip-enhanced nano-spectroscopy using a flat wavefront.
For wavefront shaping, we made a feedback loop using a simple stepwise sequential algorithm \cite{vellekoop2008}.
Specifically, 600 $\times$ 600 pixels of the SLM active area made by liquid crystals were divided into 8 $\times$ 8 segments. 
Then, each segment swept its phase from 0 to 2$\pi$ to find the optimal phase providing the strongest target signal, i.e., TEPL intensity.
We found the optimized phase mask by repeating this feedback algorithm for all the segments.\\

\noindent \threesubsection{Tip-induced strain-engineering of the nanoscale wrinkle}\\
To press and release the wrinkle apex by the Au tip (Figure 4), we gradually changed the setpoint of the shear-force feedback.
In a normal shear-force feedback condition, the distance between the tip and sample was maintained at $\sim$3 nm at the setpoint of 95 \% compared to the oscillating amplitude of the tuning fork/tip assembly in a free space.
When we gradually reduced the setpoint to 90 \%, we expected that the tip almost made contact with the wrinkle surface because we verified that the tip moved down $\sim$3 nm.
We then continuously reduced the setpoint to 50 \% and the tip gradually moved down to $\sim$10 nm lower position, compared to the tip height at the setpoint of 90 \%, while pressing the wrinkle apex.
Therefore, we estimated the pressing depth of the wrinkle to $\sim$10 nm in the experiment shown in Figure 4.
For the switching and modulation experiment in Figure 5, we programmably changed the setpoint periodically as a function of time with discrete values (90 \% and 40 \% for switching, 90 \%, 65 \%, and 40 \% for modulation).
We verified that these setpoints correspond to the pressing depths of 0, 5, and 10 nm in our experiment.\\

\noindent \threesubsection{Simulation of the tip-induced local pressure and force}\\
To quantify the tip-induced local pressure applied on the contact region between the Au tip and the wrinkle in a WSe$_2$ monolayer, we modeled our experimental conditions and calculated the pressure using a commercially available three-dimensional simulation program, the Mechanical Enterprise module from ANSYS. 
In our model, the material of the Au tip was set to gold in the program and the material properties of the WSe$_2$ monolayer, such as Young's modulus, the Poisson ratio, and density, were derived from Ref. \cite{zhang2016}.
Geometrically, the vertical position of the tip apex was located at the top of the wrinkle structure before starting the simulation.
Both ends of the WSe$_2$ monolayer wrinkle were considered to be fixed at the substrate. 
In these conditions, we calculated the local pressure applied on the wrinkle structure as a function of the pressing depth of the tip with a step of 0.5 nm.\\

\noindent \threesubsection{Simulation of modified electronic bandstructures at wrinkle}\\
The density functional theory (DFT) calculations were carried out by using the Vienna $ab initio$ simulation package (VASP) \cite{kresse1996}. 
For a WSe$_2$ monolayer, the slab model was used with the primitive lateral unit cell parameter from the experimental value \cite{schutte1987} with 15 \AA { }vacuum size. 
To describe the wrinkle observed in the experiment, we introduced 0.1 \% and 0.2 \% uniaxial strains in the monolayer. 
In the band structure calculation, the HSE06 hybrid functional including the spin-orbit coupling was used \cite{krukau2006} to study the strain effect (0.0, 0.1, 0.2 \%) on the fundamental gap. 
For the optical spectra, the Bethe-Salpeter equation was solved with the screening parameters (${\epsilon}_{\infty}^{-1}$, ${\lambda}_{TF}$, ${E}_{g}^{PBE+GW}$-${E}_{g}^{PBE}$) found from PBE and PBE+GW calculations \cite{bokdam2016, shishkin2006}. The k-mesh used in the hybrid functional and the model BSE calcultions was 12$\times$12$\times$1, while in the GW calculation for the screening parameters it was 9$\times$9$\times$1.

\medskip
\noindent \textbf{Supporting Information} \par 
\noindent Supporting Information is available from the Wiley Online Library or from the author.

\medskip
\noindent \textbf{Acknowledgements}
This work was supported by the National Research Foundation of Korea (NRF) grants (No. 2019K2A9A1A06099937, and 2020R1C1C1011301).
Computation was supported by KISTI (KSC-2019-CRE-0221).
S.H.C. and K.K.K. acknowledge the support by the Institute for Basic Science (IBS-R011-D1) and the Basic Research Program through the National Research Foundation of Korea (NRF) funded by the Ministry of Science, ICT \& Future Planning (2018R1A2B2002302).
H.S.L. acknowledges the support by the Technology Innovation Program (20005750, Commercial Development of Combustion System Control Technology for Minimizing Pollutant with Multiple Analysis) funded by the Ministry of Trade, Industry, \& Energy (MOTIE, Republic of Korea) and the Basic Science Research Program through the National Research Foundation (NRF) of Korea grant funded by the MSIT(NRF-2020R1A4A1019566).

\medskip

\noindent \textbf{Competing interests}
The authors declare no competing financial interests. 

\medskip

%
\bibliographystyle{MSP}
\bibliography{wrinkle}

\begin{thebibliography}{10}
\providecommand{\url}[1]{\texttt{#1}}
\providecommand{\urlprefix}{URL }

\bibitem{splendiani2010}
A.~Splendiani, L.~Sun, Y.~Zhang, T.~Li, J.~Kim, C.-Y. Chim, G.~Galli, F.~Wang,
\newblock \emph{Nano Lett.} \textbf{2010}, \emph{10}, 4 1271.

\bibitem{mak2010}
K.~F. Mak, C.~Lee, J.~Hone, J.~Shan, T.~F. Heinz,
\newblock \emph{Phys. Rev. Lett.} \textbf{2010}, \emph{105}, 13 136805.

\bibitem{van2018}
D.~Van~Tuan, M.~Yang, H.~Dery,
\newblock \emph{Phys. Rev. B} \textbf{2018}, \emph{98}, 12 125308.

\bibitem{kosmider2013}
K.~Ko{\'s}mider, J.~W. Gonz{\'a}lez, J.~Fern{\'a}ndez-Rossier,
\newblock \emph{Phys. Rev. B} \textbf{2013}, \emph{88}, 24 245436.

\bibitem{sie2018}
E.~J. Sie,
\newblock In \emph{Coherent Light-Matter Interactions in Monolayer
  Transition-Metal Dichalcogenides}, 37--57. Springer, \textbf{2018}.

\bibitem{mak2016}
K.~F. Mak, J.~Shan,
\newblock \emph{Nat. Photon.} \textbf{2016}, \emph{10}, 4 216.

\bibitem{park2016tmd}
K.-D. Park, O.~Khatib, V.~Kravtsov, G.~Clark, X.~Xu, M.~B. Raschke,
\newblock \emph{Nano Lett.} \textbf{2016}, \emph{16}, 4 2621.

\bibitem{park2017}
K.-D. Park, M.~B. Raschke, J.~M. Atkin, Y.~H. Lee, M.~S. Jeong,
\newblock \emph{Adv. Mater.} \textbf{2017}, \emph{29}, 7 1603601.

\bibitem{shi2013}
H.~Shi, R.~Yan, S.~Bertolazzi, J.~Brivio, B.~Gao, A.~Kis, D.~Jena, H.~G. Xing,
  L.~Huang,
\newblock \emph{ACS Nano} \textbf{2013}, \emph{7}, 2 1072.

\bibitem{miro2013}
P.~Mir{\'o}, M.~Ghorbani-Asl, T.~Heine,
\newblock \emph{Adv. Mater.} \textbf{2013}, \emph{25}, 38 5473.

\bibitem{luo2015}
S.~Luo, G.~Hao, Y.~Fan, L.~Kou, C.~He, X.~Qi, C.~Tang, J.~Li, K.~Huang,
  J.~Zhong,
\newblock \emph{Nanotechnology} \textbf{2015}, \emph{26}, 10 105705.

\bibitem{eda2012}
G.~Eda, T.~Fujita, H.~Yamaguchi, D.~Voiry, M.~Chen, M.~Chhowalla,
\newblock \emph{ACS Nano} \textbf{2012}, \emph{6}, 8 7311.

\bibitem{deng2017}
S.~Deng, E.~Gao, Z.~Xu, V.~Berry,
\newblock \emph{ACS Appl. Mater. Interfaces} \textbf{2017}, \emph{9}, 8 7812.

\bibitem{dagdeviren2020}
O.~E. Dagdeviren, O.~Acikgoz, P.~Grutter, M.~Z. Baykara,
\newblock \emph{npj 2D Mater. Appl.} \textbf{2020}, \emph{4} 30.

\bibitem{mills2016}
A.~Mills, Y.~Yu, C.~Chen, B.~Huang, L.~Cao, C.~Tao,
\newblock \emph{Appl. Phys. Lett.} \textbf{2016}, \emph{108}, 8 081601.

\bibitem{trainer2019}
D.~J. Trainer, Y.~Zhang, F.~Bobba, X.~Xi, S.-W. Hla, M.~Iavarone,
\newblock \emph{ACS Nano} \textbf{2019}, \emph{13}, 7 8284.

\bibitem{castellanos2013}
A.~Castellanos-Gomez, R.~Rold{\'a}n, E.~Cappelluti, M.~Buscema, F.~Guinea,
  H.~S. van~der Zant, G.~A. Steele,
\newblock \emph{Nano Lett.} \textbf{2013}, \emph{13}, 11 5361.

\bibitem{khan2020}
A.~R. Khan, T.~Lu, W.~Ma, Y.~Lu, Y.~Liu,
\newblock \emph{Adv. Electron. Mater.} \textbf{2020}, \emph{6}, 4 1901381.

\bibitem{lee2020}
J.~Lee, S.~J. Yun, C.~Seo, K.~Cho, T.~S. Kim, G.~H. An, K.~Kang, H.~S. Lee,
  J.~Kim,
\newblock \emph{Nano Lett.} \textbf{2020}.

\bibitem{christiansen2017}
D.~Christiansen, M.~Selig, G.~Bergh{\"a}user, R.~Schmidt, I.~Niehues,
  R.~Schneider, A.~Arora, S.~M. de~Vasconcellos, R.~Bratschitsch, E.~Malic,
  et~al.,
\newblock \emph{Phys. Rev. Lett.} \textbf{2017}, \emph{119}, 18 187402.

\bibitem{niehues2018}
I.~Niehues, R.~Schmidt, M.~Druppel, P.~Marauhn, D.~Christiansen, M.~Selig,
  G.~Berghauser, D.~Wigger, R.~Schneider, L.~Braasch, et~al.,
\newblock \emph{Nano Lett.} \textbf{2018}, \emph{18}, 3 1751.

\bibitem{koskinen2014}
P.~Koskinen, I.~Fampiou, A.~Ramasubramaniam,
\newblock \emph{Phys. Rev. Lett.} \textbf{2014}, \emph{112}, 18 186802.

\bibitem{chen2017}
M.~Chen, J.~Xia, J.~Zhou, Q.~Zeng, K.~Li, K.~Fujisawa, W.~Fu, T.~Zhang,
  J.~Zhang, Z.~Wang, Z.~Wang, X.~Jia, M.~Terrones, Z.~X. Shen, Z.~Liu, ,
  L.~Wei,
\newblock \emph{ACS Nano} \textbf{2017}, \emph{11}, 9 9191.

\bibitem{manzeli2015}
S.~Manzeli, A.~Allain, A.~Ghadimi, A.~Kis,
\newblock \emph{Nano Lett.} \textbf{2015}, \emph{15}, 8 5330.

\bibitem{hui2013}
Y.~Y. Hui, X.~Liu, W.~Jie, N.~Y. Chan, J.~Hao, Y.-T. Hsu, L.-J. Li, W.~Guo,
  S.~P. Lau,
\newblock \emph{ACS Nano} \textbf{2013}, \emph{7}, 8 7126.

\bibitem{wang2019}
K.~Wang, A.~A. Puretzky, Z.~Hu, B.~R. Srijanto, X.~Li, N.~Gupta, H.~Yu,
  M.~Tian, M.~Mahjouri-Samani, X.~Gao, A.~Oyedele, C.~M. Rouleau, G.~Eres,
  B.~I. Yakobson, M.~Yoon, K.~Xiao, D.~B. Geohegan,
\newblock \emph{Sci. Adv.} \textbf{2019}, \emph{5}, 5 eaav4028.

\bibitem{park2018dark}
K.-D. Park, T.~Jiang, G.~Clark, X.~Xu, M.~B. Raschke,
\newblock \emph{Nat. Nanotechnol.} \textbf{2018}, \emph{13}, 1 59.

\bibitem{park2019}
K.-D. Park, M.~A. May, H.~Leng, J.~Wang, J.~A. Kropp, T.~Gougousi, M.~Pelton,
  M.~B. Raschke,
\newblock \emph{Sci. Adv.} \textbf{2019}, \emph{5}, 7 eaav5931.

\bibitem{yuan2016}
M.~Yuan, L.~N. Quan, R.~Comin, G.~Walters, R.~Sabatini, O.~Voznyy, S.~Hoogland,
  Y.~Zhao, E.~M. Beauregard, P.~Kanjanaboos, Z.~Lu, D.~H. Kim, E.~H. Sargent,
\newblock \emph{Nat. Nanotechnol.} \textbf{2016}, \emph{11}, 10 872.

\bibitem{high2008}
A.~A. High, E.~E. Novitskaya, L.~V. Butov, M.~Hanson, A.~C. Gossard,
\newblock \emph{Science} \textbf{2008}, \emph{321}, 5886 229.

\bibitem{wu2014}
W.~Wu, L.~Wang, Y.~Li, F.~Zhang, L.~Lin, S.~Niu, D.~Chenet, X.~Zhang, Y.~Hao,
  T.~F. Heinz, J.~Hone, Z.~L. Wang,
\newblock \emph{Nature} \textbf{2014}, \emph{514}, 7523 470.

\bibitem{unuchek2018}
D.~Unuchek, A.~Ciarrocchi, A.~Avsar, K.~Watanabe, T.~Taniguchi, A.~Kis,
\newblock \emph{Nature} \textbf{2018}, \emph{560}, 7718 340.

\bibitem{harats2020}
M.~G. Harats, J.~N. Kirchhof, M.~Qiao, K.~Greben, K.~I. Bolotin,
\newblock \emph{Nat. Photon.} \textbf{2020}, \emph{14}, 5 324.

\bibitem{lee2020slm}
D.~Y. Lee, C.~Park, J.~Choi, M.~S. Jeong, M.~B. Raschke, K.-D. Park,
\newblock \emph{arXiv:2009.06881} \textbf{2020}.

\bibitem{he2019}
Z.~He, Z.~Han, J.~Yuan, A.~M. Sinyukov, H.~Eleuch, C.~Niu, Z.~Zhang, J.~Lou,
  J.~Hu, D.~V. Voronine, et~al.,
\newblock \emph{Sci. Adv.} \textbf{2019}, \emph{5}, 10 eaau8763.

\bibitem{feng2012}
J.~Feng, X.~Qian, C.-W. Huang, J.~Li,
\newblock \emph{Nat. Photon.} \textbf{2012}, \emph{6}, 12 866.

\bibitem{zande2013}
A.~M. Van Der~Zande, P.~Y. Huang, D.~A. Chenet, T.~C. Berkelbach, Y.~You, G.-H.
  Lee, T.~F. Heinz, D.~R. Reichman, D.~A. Muller, J.~C. Hone,
\newblock \emph{Nat. Mater.} \textbf{2013}, \emph{12}, 6 554.

\bibitem{zhang2016}
R.~Zhang, V.~Koutsos, R.~Cheung,
\newblock \emph{Appl. Phys. Lett.} \textbf{2016}, \emph{108}, 4 042104.

\bibitem{karrai1995}
K.~Karrai, R.~D. Grober,
\newblock \emph{Appl. Phys. Lett.} \textbf{1995}, \emph{66}, 14 1842.

\bibitem{kresse1996}
G.~Kresse, J.~Furthm{\"u}ller,
\newblock \emph{Phys. Rev. B} \textbf{1996}, \emph{54}, 16 11169.

\bibitem{bokdam2016}
M.~Bokdam, T.~Sander, A.~Stroppa, S.~Picozzi, D.~Sarma, C.~Franchini,
  G.~Kresse,
\newblock \emph{Sci. Rep.} \textbf{2016}, \emph{6}, 1 1.

\bibitem{shishkin2006}
M.~Shishkin, G.~Kresse,
\newblock \emph{Phys. Rev. B} \textbf{2006}, \emph{74}, 3 035101.

\bibitem{choi2017}
S.~H. Choi, S.~Boandoh, Y.~H. Lee, J.~S. Lee, J.-H. Park, S.~M. Kim, W.~Yang,
  K.~K. Kim,
\newblock \emph{ACS Appl. Mater. Interfaces} \textbf{2017}, \emph{9}, 49 43021.

\bibitem{neacsu2005}
C.~C. Neacsu, G.~Steudle, M.~B. Raschke,
\newblock \emph{Appl. Phys. B} \textbf{2005}, \emph{80}, 3 295.

\bibitem{vellekoop2008}
I.~M. Vellekoop, A.~Mosk,
\newblock \emph{Opt. Commun.} \textbf{2008}, \emph{281}, 11 3071.

\bibitem{schutte1987}
W.~Schutte, J.~De~Boer, F.~Jellinek,
\newblock \emph{J. Solid State Chem.} \textbf{1987}, \emph{70}, 2 207.

\bibitem{krukau2006}
A.~V. Krukau, O.~A. Vydrov, A.~F. Izmaylov, G.~E. Scuseria,
\newblock \emph{J. Chem. Phys.} \textbf{2006}, \emph{125}, 22 224106.

\end{thebibliography}

\end{document}